\documentclass[twocolumn,showpacs,preprintnumbers,amssymb,amsmath,article]{revtex4}
\usepackage{epsfig}
\usepackage{dcolumn}

\begin{document}

\bibliographystyle{naturemag}
\preprint{code number:}
\title{Domain wall mobility, stability and Walker breakdown in magnetic nanowires}

\author{$^1$\email{mougin@lps.u-psud.fr}A. Mougin, $^1$M. Cormier, $^1$J.P. Adam, $^{1,2}$P.J. Metaxas,  $^1$J. Ferr\'e}
 \affiliation{$^1$Laboratoire de Physique des Solides, Univ. Paris-Sud, CNRS, UMR 8502, F-91405 Orsay Cedex, France}
  \affiliation{$^2$School of Physics, MO13, University of Western Australia, 35 Stirling Highway - Crawley - WA 6009 -
Australia}

\date{\today}
%
%
\begin{abstract}

We present an analytical calculation of the velocity of a single 180$^{\circ}$ domain wall in a magnetic structure with
reduced thickness and/or lateral dimension under the combined action of an external applied magnetic field and an
electrical current. As for the case of  field-induced domain wall propagation in thick films, two motion regimes with
different mobilities are obtained, below and far above the so-called Walker field. Additionally, for the case of current
induced motion, a Walker-like current density threshold can be defined.  When the dimensions of the system become
comparable to the domain wall width, the threshold field and current density, stating the wall's internal structure
stability, are reduced by the same geometrical demagnetising factor which accounts for the confinement. This points out
the fact that the velocity dependence over an extended field/current range and the knowledge of the Walker breakdown are
mandatory to draw conclusions about the phenomenological Gilbert damping parameter tuning the magnetisation dynamics.
\end{abstract}

\maketitle

\section{Introduction}
Analysing the behaviour of a magnetic domain wall under the action of an external magnetic field is not  a new concept.
For example, magnetic bubble memories, in which information was stored in small magnetised areas delimited by domain
walls and written by displacement of the walls under magnetic fields, were designed almost 30 years
ago~\cite{Malozemoff}. Recently, an alternative approach has been proposed for reversing the magnetisation of a magnetic
cell, namely current-induced magnetisation reversal~\cite{Slonczewski96,Berger93}. In this frame, the use of
current-induced magnetic domain wall propagation, as suggested by the pioneering work of Berger~\cite{Berger78,Berger93}
is nowadays raising a growing interest. Controlling domain wall motion induced by a magnetic field and/or by
 a current flowing directly into the magnetic element, is thus a main issue in magnetism as well as a technological
challenge. Indeed, in the attempt of designing nanowires enabling fast domain wall motion, disperse results have been
published so far in regards of the field-induced motion~\cite{Konishi71,Ono99,Atkinson03,Beach05,Fukumoto05} as well as
for the current-induced one~\cite{Hayashi06,Hayashi2006,Beach06,Klaui05,Grollier,Yamaguchi,Thomas,vernier}. Moreover, the
tuning of wall damping and motion remains an open problem of magnetisation dynamics.

This article intends to give a global vision of the dependence of domain wall velocity upon magnetic field and current
density strengths in nanowires. We  analyse the field and current induced propagation, within the framework of previous
analytical studies  for the viscous regime in thick infinite samples containing a single 180$^\circ$ Bloch domain wall
driven by a magnetic field~\cite{Slonczewski72,Schryer74,Malozemoff}. Indeed, when the effect of the applied magnetic
field is sufficient to overcome pinning forces, the wall velocity $v$ depends linearly on the magnetic field $H$,
corresponding to viscous motion. Two linear velocity regimes, separated by a complex transient regime, have been
predicted to exist, below and well above a threshold field called the Walker field $H_W$~\cite{Schryer74}, delineating
the stability of the wall structure. In each linear regime, a relation between the wall mobility $m=dv/dH$, the intrinsic
Gilbert damping parameter $\alpha$~\cite{Gilbert} and the dynamic domain wall width has been
established~\cite{Slonczewski72,Schryer74}. There is in the literature a significant dispersion in the interpretation of
velocity measurements in nanowires~\cite{Ono99,Atkinson03,Beach05,Hayashi06},  particularly in relation to damping
parameters extracted from the latter. This spread, partly addressed in~\cite{Beach05}, shows how crucial the knowledge of
the Walker field is. We demonstrate that the Walker field, calculated for the case of infinite and thick samples
containing  Bloch walls, is reduced in the case of confined systems. We find also that the same geometry factor affects
the Walker-like critical current density  which limits the stability regime of the domain wall under current in
nanowires.

Up to now, the most widely used magnetic layers and devices rely on domain wall motion in systems with an in-plane
magnetic
anisotropy~\cite{Konishi71,Ono99,Atkinson03,Beach05,Hayashi06,vernier,Beach06,Klaui05,Grollier,Yamaguchi,Thomas,Hayashi2006,Fukumoto05}.
However media with  out-of-plane anisotropy are promising candidates for improving the storage density and for  designing
magnetic logic devices~\cite{Daf}. In this work, perpendicularly magnetised systems will be extensively investigated;
however the presence of an in-plane  anisotropy will not affect  the geometrical parameters relevant for the wall motion
outside of the pinning limited regime.

\section{Current and field induced propagation of a Bloch wall}

\begin{figure}[!h]
\includegraphics[width=4cm]{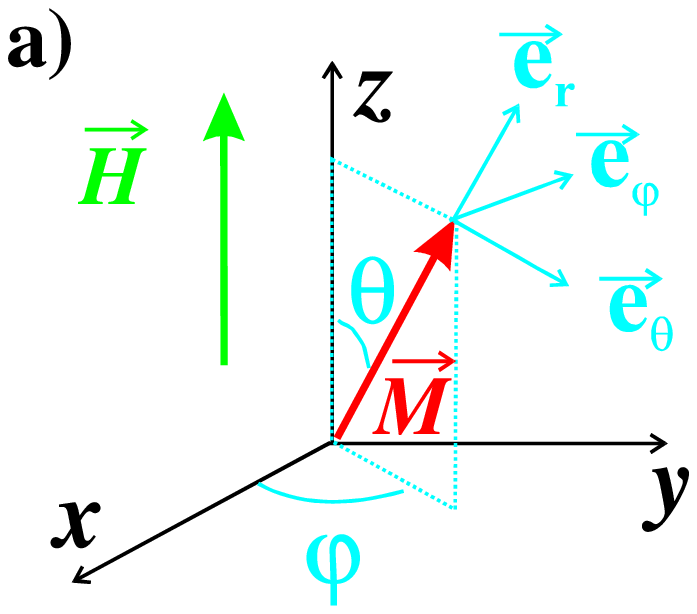}
 \includegraphics[width=5cm]{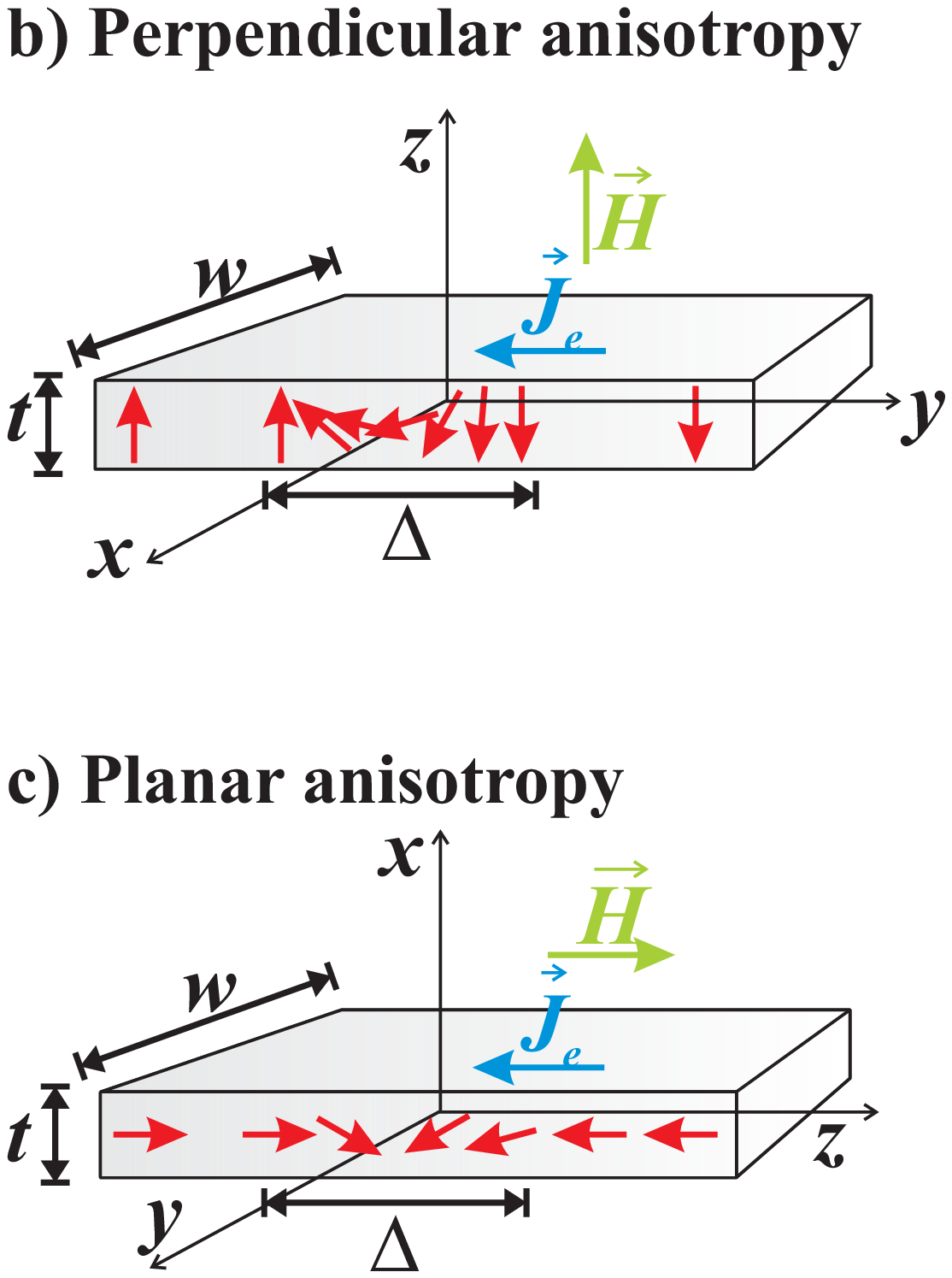}
\caption{{\bf a)}: Polar $\theta$ and azimuthal $\varphi$ angles and the associated spherical coordinate system
($\vec{e}_{r},\vec{e}_{\theta},\vec{e}_{\varphi}$) defining the orientation of the  magnetisation $\vec{M}$ relative to a
Cartesian axis system.  {\bf b)} Sketch of a 180$^\circ$ domain wall in a track of width $w$ and thickness $t$ showing a
perpendicular magnetic anisotropy.  {\bf c)} Sketch of the system of coordinates allowing to adapt the calculations to a
narrow  track with a planar anisotropy. $\Delta$ denotes the wall width  and $\vec H$ the magnetic field applied along
$z$, defined as the easy magnetisation axis of the magnetic layer in both cases. The current density $J_e$ along $y$ is
considered as positive if the electrons flow towards the positive $y$ values.}\label{geometry}
\end{figure}

We focus on the behaviour of  the local magnetisation $\vec{M}$ inside a domain wall, oriented as defined in
Fig.~(\ref{geometry}a). The length of the out-of-plane magnetised track sketched in Fig.~(\ref{geometry}b) is supposed to
be infinite, and oriented along  $y$. One first tries  to evaluate the velocity of a wall propagating towards the
positive $y$ values under the action of a magnetic field $\vec H$ applied along the easy axis of magnetisation $z$ and/or
a current density $\vec J_e$ along $y$.  We focus on structures with a 180$^\circ$ Bloch domain wall. $\theta(y)$ given
in Eq.~(\ref{Bloch}) stands as a possible wall profile;  the wall center is located at $y_0$ where $\theta = {\pi \over
2}$ and the characteristic domain wall width $\Delta$ is given in Eq.~(\ref{Delta}).  Although magnetostatic effects
within domain walls are naturally taken into account in numerical work, they are rarely discussed in analytic
treatments~\cite{donahue}. In Eq.~(\ref{Delta}) and in the following, the demagnetising factors $N_x$, $N_y$ and $N_z$
relate to the domain wall itself, $M_s$ is the saturation magnetisation, $A$ and $K_i$ are the exchange and intrinsic
anisotropy constants (see methods).
\begin{subequations}
\begin{eqnarray}
\theta (y) = 2 \arctan \biggl( e^{{{y-y_0} \over \Delta}} \biggr) \hspace{1cm} \label{Bloch}\\ \Delta = \sqrt{{A \over
{K_i + 2\pi M_s^2 \bigl( N_x \cos ^2 \varphi + N_y \sin ^2 \varphi - N_z \bigr)}}} \hspace{1cm}  \label{Delta}
\end{eqnarray}
\end{subequations}

For $\varphi = 0$ or $\pi$, one has an ``ideal'' Bloch wall with $\Delta_{\varphi=0}= \sqrt{{A \over {K_i + 2\pi M_s^2
\bigl( N_x - N_z \bigr)}}}$. This expression involves the wall demagnetising factors along those directions in which the
system has reduced dimensions: $N_x$ along the wire width and $N_z$ through the layer thickness (see methods).

 Besides the shape and size of the
sample, among the parameters affecting its mobility, the  domain wall structure is often neglected in analytic
calculations. In particular, for in-plane anisotropy, several wall types may exist depending on the film geometry:
C-shaped or N\'eel like transverse walls in infinite films~\cite{Trunk01}; N\'eel-like transverse walls or vortex walls
for a narrow wire~\cite{McMichael97}. The system of coordinates sketched on Fig.~(\ref{geometry}c) allows one to adapt,
via a permutation of the coordinates, the calculations made here for Bloch walls in out-of-plane spin configurations  to
transverse walls in in-plane spin configurations. Vortex walls, frequently observed in experiments on large nanowires,
are not considered in the present article, nor edge effects.

Magnetisation dynamics are described by Eq.~(\ref{LLG}), in which the precession of the magnetisation proposed by Landau
and Lifshitz~\cite{LL}, the dissipative effects introduced by Gilbert~\cite{Gilbert} and the usually admitted adiabatic
and non-adiabatic spin torque terms~\cite{Thiaville05,Zhang,Tatara} reflecting the  action of current, assuming  the
geometry of Fig.~(\ref{geometry}b) are merged together. Discussion on microscopic spin polarised electronic transport is
beyond the scope of this paper, in which we just discuss the macroscopic consequences of the presence of the spin-torque
terms~\cite{Thiaville05,Zhang,Tatara}.

\begin{equation} {{\partial \vec{M}} \over {\partial t}} = \gamma \vec{H}_{eff} \times \vec{M} + {\alpha \over M_s} \vec{M} \times
{{\partial \vec{M}} \over {\partial t}}- u {{\partial \vec{M}} \over {\partial y}} + {{\beta u} \over M_s} \vec{M} \times
{{\partial \vec{M}} \over {\partial y}} \label{LLG}
\end{equation}
In this expression, $\gamma$ is the gyromagnetic ratio (here  a positive quantity) and $\vec{H}_{eff}$ is the total
effective field. $\beta$ is the phenomenological non-adiabatic spin transfer parameter~\cite{Thiaville05,Zhang,Tatara}.
$u$ (Eq.~\ref{u}) has the dimension of a velocity and scales as the
 electrical current  density  $J_e$:
\begin{equation}
u = {{g J_e \mu_B P} \over {2eM_s}} \label{u}
\end{equation}
$g$ is the Land\'{e} factor, $\mu_B$ is the Bohr magnetron, $e$ is the electron charge and $P$ is the polarisation factor
of the current.

We express  the different torques acting on the magnetisation of the wall region in the
($\vec{e}_{r},\vec{e}_{\theta},\vec{e}_{\varphi}$) spherical system of coordinates (see methods).
 From the general kinetic momentum theorem,
it is  possible to relate the torque components $\Gamma_{\theta}$ and $\Gamma_{\varphi}$ and their associated precession
velocities $\dot\theta$ and $\dot\varphi$:
\begin{subequations}\label{moment_cinetique}
\begin{eqnarray}
\dot\theta = {{\partial \theta} \over {\partial t}}=-\frac{\gamma}{M_S}\Gamma_{\theta}\\
\dot\varphi = {{\partial \varphi} \over {\partial t}}=-\frac{\gamma}{M_S}\Gamma_{\varphi}
\end{eqnarray}
\end{subequations}
The first torque  to be considered (see methods)
 is that of the external field $\vec H$, namely $\vec{\Gamma}_{H}=\vec{M}\times\vec{H}$ (Eq.~\ref{H_ext}). The second one
$\vec{\Gamma}_{H_{d}}=\vec{M} \times \vec{H_d}$ (Eq.~\ref{H_d}) is that of the demagnetising field $\vec{H_d}$ within the
wall. The third torque $\vec{\Gamma}_{H_{\alpha}}=\vec{M} \times \vec{H_{\alpha}}$ (Eq.~\ref{H_a}) is that of the
equivalent damping field $\frac{-\alpha}{\gamma M_S}\vec{\frac{\partial M}{\partial t}}$.

\begin{eqnarray}
&\vec{\Gamma}_{H} =   \left(\begin{array}{c}0\\
0\\
-M_S H \sin\theta\end{array}\right)   \label{H_ext}
\end{eqnarray}
\begin{eqnarray}\vec{\Gamma}_{H_{d}}= 4\pi M_S^2  \left(\begin{array}{c} 0\\(N_y-N_x) \sin\theta \sin\varphi \cos\varphi\\
\sin\theta \cos\theta [N_z-N_y \sin^2\varphi -N_x \cos^2\varphi]  \end{array}\right)\label{H_d}\end{eqnarray}
\begin{eqnarray}\vec{\Gamma}_{H_{\alpha}}= \frac{\alpha M_S}{\gamma} \left(\begin{array}{c} 0 \\ \dot{\varphi} \sin\theta \\
 -\dot{\theta}
\end{array}\right) \label{H_a}
\end{eqnarray}

In the same spirit,  assuming that the profile at rest $\theta(y)$ given in Eq.~\ref{Bloch} is conserved under field
and/or current, which gives ${{\partial \theta} \over {\partial y}}= {{\sin\theta} \over {\Delta}}$, the adiabatic
$\vec{\Gamma}_{u}$ and the non-adiabatic $\vec{\Gamma}_{\beta}$ spin transfer torques are expressed  by
Eq.~\ref{adiabatic} and ~\ref{non-adiabatic}.
\begin{eqnarray}
\vec{\Gamma}_{u}&=&{u \over \gamma} {{\partial \vec{M}} \over {\partial \theta}}{{\partial \theta} \over {\partial y}}=
{{M_s u} \over
\gamma}{{\sin \theta} \over \Delta} \vec{e}_{\theta}\label{adiabatic}\\
\vec{\Gamma}_{\beta} &=& - {{\beta u} \over {\gamma M_s}} \vec{M} \times {{\partial \vec{M}} \over {\partial
\theta}}{{\partial \theta} \over {\partial y}}= -\beta{{M_s  u} \over \gamma}{{\sin \theta} \over \Delta}
\vec{e}_{\varphi}\label{non-adiabatic}
\end{eqnarray}

 The total polar and azimuthal torques $\Gamma_{\theta}$ and
$\Gamma_{\varphi}$   are given in Eq.~\ref{gamma_theta} and Eq.~\ref{gamma_varphi} respectively.

\begin{subequations}
\begin{eqnarray}\label{gamma_theta}
\Gamma_{\theta}=4\pi M_S^2 (N_y-N_x) \sin\theta \sin\varphi \cos\varphi \\+  \frac{\alpha M_S} {\gamma}\dot{\varphi}
\sin\theta + {{M_s u} \over \gamma}{{\sin \theta} \over \Delta}\nonumber
\end{eqnarray}
\begin{eqnarray}\label{gamma_varphi}
\Gamma_{\varphi}&=&-M_S H \sin\theta-
 \frac{\alpha M_S} {\gamma}  \dot\theta -\beta{{M_s  u} \over \gamma}{{\sin \theta} \over
\Delta}\\&+&4\pi M_S^2 \sin\theta \cos\theta [N_z-N_y \sin^2\varphi -N_x \cos^2\varphi]\nonumber
\end{eqnarray}
\end{subequations}

\section{Wall velocity and stability in systems of reduced dimensions}

For a low  applied field or current density, in ultrathin perpendicular magnetised systems, pinning dominates and domain
walls move following the creep law~\cite{Lemerle98}. In this non-linear regime, the motion is thermally activated and
proceeds by discrete jumps from one pinned configuration to the next. In the following, we assume that pinning effects
are overcome and consider only viscous propagation.  We first consider the steady domain wall motion, whose signature is
a wall moving with a time independent  azimuthal angle $\varphi$: $\dot\varphi =0$ which implies that $\Gamma_{\varphi}$
(Eq.~\ref{gamma_varphi}) must vanish.  Writing $\Gamma_{\varphi}=0$ for $\theta=\pi/2$ (at the wall center where
$\Gamma_{\varphi}$ has an extremum) gives:

 \begin{eqnarray} \sin2\varphi = \frac{H+(\beta-\alpha)\frac{u}{\gamma\Delta}}{2\pi \alpha M_s (N_y - N_x)}\label{sin2f}\\
 {\rm  only \ valid \  if } \quad \bigl|H+(\beta-\alpha)\frac{u}{\gamma\Delta}\bigr| \leq H_W\label{stability}\\
{\rm where \  we \ define \ the \ adapted \ Walker \ field \ by} \nonumber\\
 H_W=2\pi \alpha M_s \bigl|N_y - N_x
\bigr|\label{H_W}\end{eqnarray}

Eq.~\ref{stability} sets a limit on the combined strength of the applied field and current density for  steady motion to
occur.  $\left| H_{ext} + (\beta - \alpha) {u \over {\gamma\Delta}} \right|$ determines the Walker-like stability
condition for the wall motion as a whole. This limit condition is strictly equivalent to the Walker breakdown condition,
in the case where only an external magnetic field is applied. Above the Walker breakdown, the stability condition
$\dot\varphi = 0$ is broken and the motion is no longer steady. There, the torque of the external field/current dominates
those of the demagnetising (Eq.~\ref{H_d}) and damping (Eq.~\ref{H_a}) fields.  It is the $\varphi$ torque component
which fixes the steady regime limit for a wall but the underlying quantity governing the wall's stability is the
magnetostatic field inside the wall, making its motion geometry dependant. As a direct illustration, in the steady regime
($\dot \varphi$=0) and without current, the only non zero component of the damping torque is the  $\varphi$ component,
which  is proportional to the rate of change of $\theta$ and depends only on the demagnetising field. As compared to the
expression obtained by Schryer and Walker~\cite{Schryer74}, our expression of $H_W$ (Eq.~\ref{H_W}) includes a geometry
factor $\bigl|N_y - N_x \bigr|$. For a zero applied field ($H = 0$) one similarly gets a Walker breakdown current density
$J_W$ (Eq.~\ref{walker_current}). The same geometrical reduction factor enters the critical Walker-like current density
defining the stability condition of the wall.

\begin{equation}\label{walker_current}
J_W = {{4 \pi \alpha eM_s^2} \over {g\mu_B P}} {{\gamma \Delta} \over {\left| \beta - \alpha \right|}} |N_y - N_x|
\end{equation}

The wall velocity is related to the way the spin aligns along the external field, i.e. to the rate of change of $\theta$
and therefore comes out from the resulting torque $\Gamma_{\theta}$ given in Eq.~\ref{gamma_theta}.  Making use of
Eq.~\ref{moment_cinetique}, in the steady regime ($\dot\varphi=0$), $\dot\theta$ is simply $ -\gamma \times 4\pi M_S
(N_y-N_x) \sin\theta \sin\varphi \cos\varphi - u \frac{\sin\theta}{\Delta}$.   The velocity $v$ of any spin embedded in a
Bloch wall is related to the characteristic width by: $v= -{\Delta \over \sin\theta} {\dot\theta} $.  Choosing again
$\theta=\pi/2$ and using Eq.~\ref{sin2f}, we obtain:

\begin{equation}\label{v_below}
v_{steady} = {{\gamma \Delta} \over {\alpha}} \left[H+ (\beta-\alpha)\frac{u}{\gamma\Delta}\right]+u = {{\gamma \Delta}
\over {\alpha}} \left(H+\frac{\beta u}{\gamma\Delta}\right)
\end{equation}

This expression is formally equivalent to the one obtained in the Walker theory~\cite{Schryer74}, the magnetic field
being here shifted by a spin transfer equivalent field $\beta u / \gamma \Delta$. This extra term is linked to the
non-adiabatic spin transfer, which is the only spin transfer component efficient for steady motion. As shown for in-plane
magnetised samples, the adiabatic component distorts the wall structure but does not induce a stationary
motion~\cite{Thiaville05,Tatara}. It is worth noting that the ``$\beta$ term'' does not appear explicitly in the
$\Gamma_\theta$ torque component. It only appears implicitly via its action on the $\varphi$ precession.

If $|H+(\beta-\alpha)\frac{u}{\gamma\Delta}| \gg 2\pi \alpha M_s \bigl|N_y - N_x \bigr|$, we assume that the precession
of the magnetisation  occurs at a constant angular speed so that the average value over time of $\sin 2 \varphi (y_0,t)$
can be set to zero. Thus, averaging Eq.~\ref{gamma_theta} and Eq.~\ref{gamma_varphi} over a period of the precession of
$\varphi$ and making use of Eq.~\ref{moment_cinetique}, one gets the average velocity far above the Walker breakdown,
simply given by Eq.~\ref{v_above}:
\begin{eqnarray}
\overline{v} = \gamma\Delta {\alpha \over {1+\alpha^2}} \left( H + {{\beta u} \over {\gamma\Delta}} \right) + {{u} \over
{1+\alpha^2}}\label{v_above}
\end{eqnarray}

\begin{figure}
\includegraphics[width=8cm]{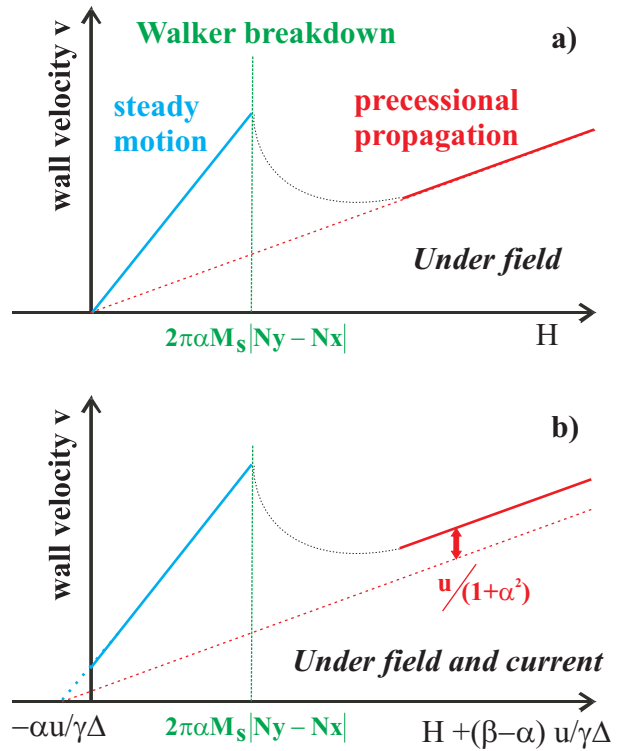}
\caption{Sketch of a 180$^\circ$ domain wall's velocity as a function  of (a) an external magnetic field $H$ and (b) a
driving term combining a field and a current ($u$ relates to the current density) in a nanowire. This cartoon indicates
the two linear regimes of velocity, below and far above the Walker breakdown. The dotted line in the transient non-linear
regime is a guide for the eyes.}\label{vitesse}
\end{figure}

This regime is similar to the usual high field one described for a 180$^{\circ}$ Bloch
wall~\cite{Slonczewski72,Schryer74,Malozemoff}, i.e. the average velocity is linear in the field, following an initial
drop in the mobility  at the breakdown.   The current action is thus i) equivalent to an additional magnetic field $\beta
u / \gamma \Delta$ applied along the anisotropy direction of the system~\cite{Thiaville05,Tatara} and ii) an extra
velocity linked to the adiabatic spin transfer, that is $u / (1 + \alpha^2)$, far above the Walker breakdown. Both linear
velocity regimes are depicted in Fig.~\ref{vitesse}, under sole action  of a driving field (a) and under the combined
action of a field and a current (b).

 In a one dimensional statement of wall motion, without current, a qualitative understanding of the
damping/demagnetising interrelation is easy to provide. When a drive field is applied along the anisotropy direction,
$\vec{M}$ starts a $\varphi$ precession movement, and tilts away with respect to its equilibrium orientation at rest,
according to the external field torque given in Eq.~\ref{H_ext}. The  out of the $xz$ wall plane component of the
magnetisation creates magnetic charges. The $\theta$ component of the torque resulting from the induced magnetostatic
field (Eq.~\ref{H_d}) describes the resulting additional precession of the spin around $\vec{H}_d$. Damping and
demagnetising torques then rapidly ``freeze'' the precession by cancelling the Zeeman torque.  At the Walker field, the
equilibrium that exists between the damping/demagnetising torque and the drive field torque keeping $\varphi$ constant is
broken and the wall exhibits its maximum velocity for $\varphi=\pi/4$ in the linear region.  Above the Walker breakdown,
 the azimuthal angle precesses around the external field. The motion is
oscillatory, due to the periodic nature of the torque term with respect to $\varphi$. In the case where there would be no
damping, the wall would move back and forth about its initial position. In the presence of damping, the forward
displacement is larger  than the backward one. This results in a positive average  displacement during one period and
hence in a non zero average velocity.

\section{Discussion}
\begin{figure}[!h]
 \includegraphics[width=6cm]{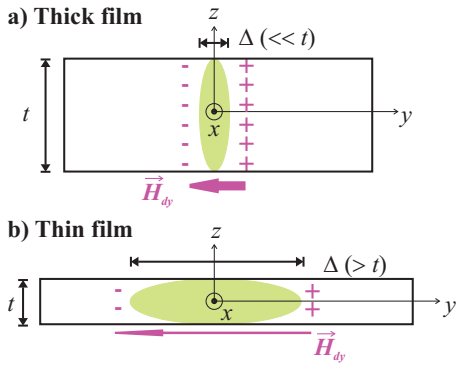}
\caption{Sketch of the $yz$ plane of a 180$^\circ$ domain wall with magnetic charges and the associated $y$ component of
the demagnetising field ($\vec{H}_{d_y}$) through the wall plane for different thicknesses. $\vec{H}_{d_y}$  is much
smaller in case b) than in case a).}\label{demagnetising_field}
\end{figure}

In contrast to the case of a thick infinite film,  one must consider all components of the wall's internal demagnetising
field to describe the Walker breakdown of the wall under the action of either a magnetic field or a spin polarised
electrical current when the system's dimensions are comparable to, or less than,  the domain wall width. For a thick and
infinite sample, the only non-zero wall demagnetising coefficient is the one along $y$ since the magnetic charges along
$z$ (thickness) and $x$ (width) are repelled towards infinity (Fig.~\ref{demagnetising_field}). Said another way, in that
geometry, the demagnetising field has only one significant component which is perpendicular to the wall plane  and the
additional torque related to $\vec{H}_{d_y}$ (through the wall width) is the only one that is efficient in aligning the
spin and pushing the wall forward. For a track of reduced dimensions and/or an ultrathin film ($t\ll\Delta$,
$w\ll\Delta$), the coefficients $N_{x}$ and $N_{z}$ are not zero anymore and reflect the aspect ratio $w/\Delta$ and
$t/\Delta$ (Fig.~\ref{demagnetising_field}). Since $N_x + N_y + N_z = 1$, the demagnetising field $\vec{H}_{d_y}$
perpendicular to the wall and its associated torque will be reduced with respect to the infinite thick film case. Time
resolved Kerr microscopy experiments performed on continuous Pt~(4.5~nm)~/ Co~($0.4~nm<t<0.9~nm$)~/ Pt~(3.5~nm) layers
grown by sputtering with a perpendicular to the plane magnetic anisotropy support the present conclusions~\cite{Metaxas}.
In the case of a nanowire of width $w$=200~nm patterned from such a Pt(4.5nm)/Co (0.5~nm)/Pt(3.5nm) film, we model the 1D
wall cross-section
 as an ellipse which allows use of the standard expression  $N_{y}\simeq t/(t+\Delta)$ and $N_{x}\simeq
 t/(t+w)$ \cite{Osborn45}. This leads to $N_x \simeq$ 0.0025,  $N_y \simeq$  0.0775 and $N_z \simeq$ 0.92. The Walker threshold
 field $H_W$ of the nanowire is then about 7.5\% of that of the corresponding thick and infinite layer.

It is in principle possible to evaluate the damping constant $\alpha$ from domain wall velocity experiments. Under field,
this has been tentatively done using in-plane magnetised samples ~\cite{Konishi71,Ono99,Atkinson03,Beach05} but rarely
for out-of-plane magnetised systems~\cite{Tarasenko98,Kirilyuk93}, in which the creep regime  often extends over a large
field range~\cite{Lemerle98} and prevents the observation of the linear regimes. The velocity dependence over an extended
field/current range and the knowledge of the Walker breakdown are mandatory to draw conclusions about the
phenomenological Gilbert damping parameter $\alpha$. Experimentally, the first unambiguous evidence of the initial linear
regime has been obtained in the case of large area NiFe films (with in-plane magnetisation) of thickness ranging between
310 and 3000~\AA\ ~\cite{Konishi71}. The authors verified  the Bloch nature of the domain wall (above 1000~\AA) and then
observed an increase of the mobility with the thickness accounted by the domain wall width evolution; for smaller
thicknesses and/or samples containing N\'eel walls, either it was not possible to observe  the viscous regime or the
thickness dependance was inverse. In the case of a few hundreds \AA\ thick NiFe nanowires of sub-micrometric width,
$v(H)$ experiments have been reported~\cite{Ono99,Atkinson03,Beach05,Hayashi06}. However, the two successive regimes and
the Walker threshold have been evidenced only recently in in-plane systems~\cite{Beach05,Hayashi06}. For current induced
motion,  the Walker breakdown was also evidenced recently~\cite{Thomas,Beach06,Hayashi2006}.  Satisfactorily,
in~\cite{Hayashi2006}, the damping constant used to account for the experimental value of the Walker field is compatible
with our statement about $H_W$.

\section{Conclusion}

In summary, in the frame of designing devices based on domain wall motion, understanding of the Walker breakdown is
critical.  Above this breakdown, the wall structure is unstable causing the instantaneous wall velocity to oscillate and
the average velocity to drop abruptly.  We have demonstrated that wall confinement effects must be taken into account. In
nanostructures, magnetostatic effects within the wall lead to a restriction of the stability region relative to thick and
infinite films (Eq.~\ref{H_W});  the domain wall width parameter $\Delta$ (Eq.~\ref{Delta}) is also modified. The domain
wall structure tuning  is also a tricky problem and no consensus has been achieved yet concerning the evaluation of
$\alpha$. For current induced domain wall motion, dispersed results have been obtained so far, partly because of the
distortion of the wall, especially in in-plane systems. Therefore, the choice of perpendicularly magnetised ultrathin
layers, whose wall structure is known and in which the non adiabatic contribution should be large~\cite{Tatara}, are
probably good candidates for a detailed understanding of the underlying mechanisms of wall
motion and damping as well as for reliability of devices.\\

\section{Methods}

\subsection{Wall's parameters} In the case of a uniformly magnetised ellipso\"id, the internal magnetostatic field
$\vec{H}_d$ is proportional to the magnetisation according to
$\vec {H}_{d}=-4\pi\left(\begin{array}{ccc}N_x&0&0\\0&N_y&0\\0&0&N_z\\
\end{array}\right) \vec M$.  The demagnetising factors $N_x$, $N_y$ and $N_z$  along the $x$, $y$ and $z$ directions respectively
verify $N_x + N_y + N_z = 1$.  Magnetostatic demagnetising factors are complex to calculate in
general~\cite{Aharoni98,Malozemoff} but they can be evaluated for simple shapes~\cite{Osborn45,Aharoni98}. \\
In the case where an out-of-plane uniform magnetic layer (and not a domain wall) is considered, an effective anisotropy
constant is measured: $K=K_{i} - 2\pi M_s^2$, that takes into account the intrinsic anisotropy constant $K_{i}$ and the
demagnetising field of the out-of-plane saturated layer
($N_x^{layer}=N_y^{layer}=0$ and $N_z^{layer}$= 1). \\
In the case of a pure Bloch wall ($\varphi=0$), one can compare $\Delta_{\varphi=0}$  with the domain wall width
parameter of a Bloch wall at rest in an infinite and thick layer~\cite{Schryer74}, that is
$\Delta_{Bloch}^{thick}=\sqrt{A/K_i}$. For an
 infinite but thin layer, the Bloch domain wall width becomes $\Delta_{Bloch}^{thin}=\sqrt{A/(K_i- 2\pi M_s^2
N_z)}=\sqrt{A/(K+ 2\pi M_s^2 N_y)}$.

\subsection{Torque calculations} The  torques of the exchange and anisotropy fields compensate each other and therefore
are not included in the torque calculations. They define the wall profile at rest, that is considered as conserved when
investigating the dynamics. \\
The active torques given in Eq.~\ref{H_ext},~\ref{H_d} and ~\ref{H_a} are obtained using the transfer matrix $\vec
\Gamma_{r,\theta,\varphi}= \left(\begin{array}{ccc}
 \sin\theta \cos\varphi&\sin\theta \sin\varphi&\cos\theta\\
 \cos\theta \cos\varphi&\cos\theta \sin\varphi& -\sin\theta\\
 -\sin\varphi &\cos\varphi&0  \end{array}\right)\vec \Gamma_{x,y,z}$.

\begin{acknowledgments}
R.L. Stamps and A. Thiaville are acknowledged for stimulating discussions and a critical reading of the manuscript.
J.P.A., J.F. and A.M. acknowledge support by FAST-Egide. Support by the french ``ACI PARCOUR'' is acknowledged. The stay
at Orsay of P.M. is supported by ``Marie-Curie Orsay Early Stage Training Site on Emergent Phenomena in Condensed Matter
Physics'' and FAST. P.M. also acknowledges the Australian Government.

\end{acknowledgments}

\bibliography{references_mougin}

\end{document}